\documentstyle[aps,12pt]{revtex}

\begin{document}
\title{Bounds for state-dependent quantum cloning}
\author{Yong-Jian Han, Yong-Sheng Zhang\thanks{%
Electronic address: yshzhang@ustc.edu.cn}, Guang-Can Guo\thanks{%
Electronic address: gcguo@ustc.edu.cn}}
\address{Key Laboratory of Quantum Information, University of Science and Technology\\
of China, CAS, Hefei 230026, People's Republic of China\bigskip \bigskip }
\maketitle

\begin{abstract}
\baselineskip24pt Due to the no-cloning theorem, the unknown quantum state
can only be cloned approximately or exactly with some probability. There are
two types of cloners: universal and state-dependent cloner. The optimal
universal cloner has been found and could be viewed as a special
state-dependent quantum cloner which has no information about the states. In
this paper, we investigate the state-dependent cloning when the state-set
contains more than two states. We get some bounds of the global fidelity for
these processes. This method is not dependent on the number of the states
contained in the state-set. It is also independent of the numbers of copying.

PACS number(s): 03.67.-a, 03.65.Ta, 89.70.+c\medskip 
\end{abstract}

\baselineskip20pt

\section{Introduction}

The no-cloning theorem is one of the most important characters of quantum
information, which is different from classical information. On the basis of
superposition principle, Wootters and Zurek\cite{woo82} pointed out that it
is impossible to find a way to copy an arbitrary unknown state perfectly.
They introduced a cloner which is named Wootters-Zurek Copying Machine (W-Z
CM). This machine can copy orthogonal state perfectly, but copy the
superposition states badly. Since determinately perfect copying is
impossible, the approximate cloning is necessary. Bu\v {z}ek and Hillery\cite
{hb96} have first shown that the universal cloner is possible and introduced
a copying machine which is called Bu\v {z}ek-Hillery Copying Machine (B-H
CM). This machine is deterministic and does not need any information about
the state to be cloned. It can copy every state equally well. Then it has
been proved\cite{gm97,bem98} that the B-H CM is the optimal cloning machine
for universal cloning, that is, this machine attains the largest local
fidelity. There is another kind of cloner which is named state-dependent
cloner. It needs some information about the cloning state. There are three
types of this kind of cloner: deterministic, probabilistic and hybrid
cloners. Probabilistic cloner has been introduced by Duan and Guo\cite
{dg97,98}. They found the states could be cloned perfectly with some
probability less than 1,when the states are linearly independent. The
deterministic state-dependent cloner was first investigated by Bru\ss\ {\it %
et al.}\cite{brub98} and it was solved completely when the state-set
contains only two states which have equally {\it a priori }probability. Then
Chefles and Barnett\cite{chef99} generalized this problem to the two states
which have different {\it a priori }probability and the global fidelity is
used to measure of the cloning process instead of the local fidelity. They
gave the optimal strategy to make the global fidelity maximal and found this
fidelity is larger than the universal cloner. Several months latter, Chefles
and Barnett\cite{chef99} hybridized the former two types of cloners to get
the hybridized cloner. So the former two cloners can be viewed as a special
case of it.

When the state-set contains only two states, there is a analytic solution of
the optimal strategy. Unfortunately, when the number of the states is more
than two, there is no analytic solution for this problem. Before the exact
solution of 2-state-dependent quantum cloner was found, some scientists had
already began to study the bound of these processes. The original work was
proposed by Hillery and Bu\v {z}ek\cite{hb97}, they derived a lower bound
for the amount of the noise introduced by quantum cloning process. More
recently, Rastegin\cite{r0108,0111} gave another lower bound for the noise
by a new method. Since solving this problem exactly is impossible when
state-set contains more than two states, it is necessary to find the bound
of the global fidelity of the quantum cloning process. By the way, when the
number of the states is increasing, these states are no longer linearly
independent. So the Duan-Guo cloning machine does not work. Even though the
bound can not tell us what we can do, it can only tell us what we can not
attain. In this paper, some bounds on the multi-state-dependent quantum
cloning process are given. We study the three-state-dependent (that is, the
state-set contains three states) quantum copying more carefully and
generalize the method to the multi-state-dependent cloning process. In
Section II, some necessary lemmas are introduced. In Section III, the upper
bound of the global fidelity of 3-state-dependent quantum cloning process is
given. In Section IV, some upper bounds for n-state-dependent quantum
copying process are introduced. The conclusion is given in Section V.

\section{Necessary premise}

Consider a set of $n$ nonorthogonal quantum states $\left| \psi
_i\right\rangle (1\leq i\leq n)$. If there are $M$ quantum systems, they are
prepared in the same unknown quantum state $\left| \phi \right\rangle $
which is taken from the given set. The task is to find an optimal process to
get $N>M$ identical approximate cloning states from the $M$ initial states $%
\left| \phi \right\rangle $. This process is a symmetric cloning, which can
be denoted by $M\rightarrow N$. The optimal process means the global
fidelity is maximal. The global fidelity is defined as follows: 
\begin{eqnarray}
F_{MN} &=&\sum\limits_{j=1}^n\eta _j\left| \left\langle \Psi _j^N\right|
\left. \Phi _j\right\rangle \right| ^2  \eqnum{1} \\
&=&\sum\limits_{j=1}^n\eta _j\left| \left\langle \Psi _j^N\right| U\left|
\Psi _j^M\right\rangle \otimes \left| 0\right\rangle \right| ^2,  \nonumber
\end{eqnarray}
where $\left| \Phi _j\right\rangle $ denotes the actual $N$ copies of cloned
state of $\left| \psi _j\right\rangle $, $\left| \Psi _j^N\right\rangle $
denotes the exact $N$ copies of cloned state of $\left| \psi _j\right\rangle 
$ which is a N-fold tensor and $\eta _j$ stands for the {\it a priori}
probability of the state $\left| \psi _j^M\right\rangle $, and $\left|
0\right\rangle $ denotes the $N-M$ blank copies.

The case of the state-set only containing two states has already been solved
by Bru\ss\ {\it et al.}\cite{brub98} and Chefles and Barnett \cite{chef99}.
When they derived the optimal strategy, the following fact is crucial: the
optimal outputs $\left| \Phi _{\pm }\right\rangle $ lie in the subspace
spanned by the exact clones $\left| \psi _{\pm }^N\right\rangle $. It is
also held when the state-set has more than two states. In fact, we have the
following lemma:

{\it Lemma 1. }For any state set $S=\{\left| \psi _1\right\rangle ,\left|
\psi _2\right\rangle \cdots \left| \psi _n\right\rangle \}$ (assume the {\it %
a priori }probability of each state are $\frac 1n$) and for the quantum
cloning process $M\rightarrow N,$ the optimal outputs $\{\left| \Phi
_1\right\rangle ,\left| \Phi _2\right\rangle \cdots \left| \Phi
_n\right\rangle \}$ lie in the subspace spanned by the exact clones $%
\{\left| \psi _1^N\right\rangle ,\left| \psi _2^N\right\rangle \cdots \left|
\psi _n^N\right\rangle \}.$

{\it Proof.}

This proof is following the method introduced by Bru\ss\ {\it et al.}\cite
{brub98}{\it .}

At the beginning of this proof, we can define a matrix $\Xi $ of the
state-set as 
\begin{equation}
\Xi =\left( 
\begin{array}{llll}
\left\langle \psi _1\right| \left. \psi _1\right\rangle & \left\langle \psi
_1\right| \left. \psi _2\right\rangle & \cdots & \left\langle \psi _1\right|
\left. \psi _n\right\rangle \\ 
\left\langle \psi _2\right| \left. \psi _1\right\rangle & \left\langle \psi
_2\right| \left. \psi _2\right\rangle & \cdots & \vdots \\ 
\vdots & \vdots & \ddots & \vdots \\ 
\left\langle \psi _n\right| \left. \psi _1\right\rangle & \cdots & \cdots & 
\left\langle \psi _n\right| \left. \psi _n\right\rangle
\end{array}
\right) .  \eqnum{2}
\end{equation}
This matrix is necessary in the following and the definition shows that it
is a Hermite matrix.

Suppose that the optimal outputs have the other components which do not lie
in the subspace spanned by the exact clones. Then the optimal outputs can be
written as 
\begin{eqnarray}
U\left| \psi _1^M\right\rangle \otimes \left| 0\right\rangle &=&\left| \Phi
_1\right\rangle =a_{11}\left| \psi _1^N\right\rangle +a_{12}\left| \psi
_2^N\right\rangle +\cdots +a_{1n}\left| \psi _n^N\right\rangle +b_1\left|
\Gamma _1\right\rangle  \eqnum{3} \\
U\left| \psi _2^M\right\rangle \otimes \left| 0\right\rangle &=&\left| \Phi
_2\right\rangle =a_{21}\left| \psi _1^N\right\rangle +a_{22}\left| \psi
_2^N\right\rangle +\cdots +a_{2n}\left| \psi _n^N\right\rangle +b_2\left|
\Gamma _2\right\rangle  \nonumber \\
&&\vdots  \nonumber \\
U\left| \psi _n^M\right\rangle \otimes \left| 0\right\rangle &=&\left| \Phi
_n\right\rangle =a_{n1}\left| \psi _1^N\right\rangle +a_{n2}\left| \psi
_2^N\right\rangle +\cdots +a_{nn}\left| \psi _n^N\right\rangle +b_n\left|
\Gamma _n\right\rangle ,  \nonumber
\end{eqnarray}
where the vectors $\left| \Gamma _1\right\rangle ,\left| \Gamma
_2\right\rangle \cdots \left| \Gamma _n\right\rangle $ are normalized and
orthogonal to the subspace spanned by the exact clones, and $\left|
0\right\rangle $ denotes the $N-M$ blank copies. Since the transformation is
unitary, the following constraints must be held. 
\begin{equation}
\Omega _{ij}^1=%
\mathop{\rm Re}%
[\sum\limits_{k,l}\Xi _{kl}^Na_{ik}^{*}a_{jl}+b_i^{*}b_j\left\langle \Gamma
_i\right| \left. \Gamma _j\right\rangle -\Xi _{ij}^M]=0,  \eqnum{4}
\end{equation}
\[
\Omega _{ij}^2=%
\mathop{\rm Im}%
[\sum\limits_{k,l}\Xi _{kl}^Na_{ik}^{*}a_{jl}+b_i^{*}b_j\left\langle \Gamma
_i\right| \left. \Gamma _j\right\rangle -\Xi _{ij}^M]=0, 
\]
where $\Xi _{kl}^N=\left\langle \psi _k^N\right| \left. \psi
_l^N\right\rangle =(\Xi _{kl})^N$ , $N$ denotes the number of copies and $M$
denotes the number of initial identical states. Particularly, when $i=j$,
there is the following constraints 
\begin{equation}
\Omega _{ii}=\sum\limits_{k,l}\Xi _{kl}^Na_{ik}^{*}a_{il}+\left| b_i\right|
^2-1=0.  \eqnum{5}
\end{equation}

The global fidelity is 
\begin{eqnarray}
F_{MN} &=&\frac 1n\sum\limits_{i=1}^n\left| \sum_{j=1}^n\Xi
_{ij}^Na_{ij}\right| ^2  \eqnum{6} \\
&=&\frac 1n\sum\limits_{i=1}^n[\sum\limits_{k,l}(\Xi _{ik}^N\Xi
_{il}^{*N}-\Xi _{lk}^N)a_{il}^{*}a_{ik}+\sum\limits_{k,l}\Xi
_{lk}^Na_{il}^{*}a_{ik}]  \nonumber \\
&=&\frac 1n\sum\limits_{i=1}^n(1-\left| b_i\right| ^2)+\frac 1n%
\sum\limits_{i=1}^n[\sum\limits_{k,l}(\Xi _{ik}^N\Xi _{il}^{*N}-\Xi
_{lk}^N)a_{il}^{*}a_{ik}.  \nonumber
\end{eqnarray}

The constraints (5) have already been used. Now we can use the Lagrange
multipliers for the other constraints and get these equations 
\begin{equation}
\frac{\partial F_{MN}}{\partial \left| a_{ij}\right| }+\sum_{k,l,\sigma
}\lambda _{kl}^\sigma \frac{\partial \Omega _{kl}^\sigma }{\partial \left|
a_{ij}\right| }=0,  \eqnum{7}
\end{equation}
\begin{equation}
\frac{\partial F_{MN}}{\partial \left| b_i\right| }+\sum_{k,l,\sigma
}\lambda _{kl}^\sigma \frac{\partial \Omega _{kl}^\sigma }{\partial \left|
b_i\right| }=0,  \eqnum{8}
\end{equation}
\begin{equation}
\frac{\partial F_{MN}}{\partial \left| \left\langle \Gamma _i\right. \left|
\Gamma _j\right\rangle \right| }+\sum_{k,l,\sigma }\lambda _{kl}^\sigma 
\frac{\partial \Omega _{kl}^\sigma }{\partial \left| \left\langle \Gamma
_i\right. \left| \Gamma _j\right\rangle \right| }=0  \eqnum{9}
\end{equation}
etc, Where $\Omega _{kl}^\sigma $ denotes the constraints, and the Lagrange
multipliers are $\lambda _{kl}^\sigma $. Since the constraint when $k=l$ has
been used before, the index $k$ and $l$ in all of the equations must satisfy 
$k\neq l$. We suppose that $b_i=\left| b_i\right| e^{i\delta _i}$ and $%
\left\langle \Gamma _k\right| \Gamma _i\rangle =\left| \left\langle \Gamma
_k\right| \Gamma _i\rangle \right| e^{i\delta _{ki}}$. Due to the equality
of $\Omega _{kl}^\sigma $ and $\Omega _{lk}^\sigma $, we can only consider
one of them. Then Eq. (8) and Eq. (9) can be written as 
\begin{equation}
-\frac 2n\left| b_i\right| +\sum_{k=1}^n\lambda _{ki}^1%
\mathop{\rm Re}%
(e^{i(\delta _i-\delta _k)}\left| b_k\right| \left\langle \Gamma _k\right|
\Gamma _i\rangle )+\sum_{k=1}^n\lambda _{ki}^2%
\mathop{\rm Im}%
(e^{i(\delta _i-\delta _k)}\left| b_k\right| \left\langle \Gamma _k\right|
\Gamma _i\rangle )=0,  \eqnum{10}
\end{equation}
\begin{equation}
\lambda _{ji}^1%
\mathop{\rm Re}%
(e^{i(\delta _i-\delta _j+\delta _{ji})}\left| b_i\right| \left| b_j\right|
)+\lambda _{ji}^2%
\mathop{\rm Im}%
(e^{i(\delta _i-\delta _j+\delta _{ji})}\left| b_i\right| \left| b_j\right|
)=0.  \eqnum{11}
\end{equation}
Let us multiply Eq. (10) by $\left| b_i\right| ,$ we get

\begin{equation}
-\frac 2n\left| b_i\right| ^2+\sum_{k=1}^n\lambda _{ki}^1%
\mathop{\rm Re}%
(e^{i(\delta _i-\delta _k)}\left| b_k\right| \left| b_i\right| \left\langle
\Gamma _k\right| \Gamma _i\rangle )+\sum_{k=1}^n\lambda _{ki}^2%
\mathop{\rm Im}%
(e^{i(\delta _i-\delta _k)}\left| b_k\right| \left| b_i\right| \left\langle
\Gamma _k\right| \Gamma _i\rangle )=0.  \eqnum{12}
\end{equation}
After multiplying Eq. (11) by $\left| \left\langle \Gamma _j\right| \Gamma
_i\rangle \right| $, we find

\[
\lambda _{ji}^1%
\mathop{\rm Re}%
(e^{i(\delta _i-\delta _j+\delta _{ji})}\left| b_i\right| \left| b_j\right|
\left| \left\langle \Gamma _j\right| \Gamma _i\rangle \right| )+\lambda
_{ji}^2%
\mathop{\rm Im}%
(e^{i(\delta _i-\delta _j+\delta _{ji})}\left| b_i\right| \left| b_j\right|
\left| \left\langle \Gamma _j\right| \Gamma _i\rangle \right| )=0, 
\]
that is, 
\[
\lambda _{ji}^1%
\mathop{\rm Re}%
(e^{i(\delta _i-\delta _j)}\left| b_i\right| \left| b_j\right| \left\langle
\Gamma _j\right| \Gamma _i\rangle )+\lambda _{ji}^2%
\mathop{\rm Im}%
(e^{i(\delta _i-\delta _j)}\left| b_i\right| \left| b_j\right| \left\langle
\Gamma _j\right| \Gamma _i\rangle )=0. 
\]
Then we sum them over the subscript $j$ from$1$ to $n$ and get

\begin{equation}
\sum_{j=1}^n[\lambda _{ji}^1%
\mathop{\rm Re}%
(e^{i(\delta _i-\delta _j)}\left| b_i\right| \left| b_j\right| \left\langle
\Gamma _j\right| \Gamma _i\rangle )+\lambda _{ji}^2%
\mathop{\rm Im}%
(e^{i(\delta _i-\delta _j)}\left| b_i\right| \left| b_j\right| \left\langle
\Gamma _j\right| \Gamma _i\rangle )]=0.  \eqnum{13}
\end{equation}

Substituting Eq.(13) into Eq. (12) and changing the subscript $j$ to $k$, we
can find that $\left| b_i\right| ^2=0$, that is, $\left| b_i\right| =0$.
This is the end of the proof.

Note that the lemma is also held when the {\it a priori }probability is not
equal for all of the states. The proof is the same as before. What we need
to do is to change the {\it a priori }probability $\frac 1n$ by the new {\it %
a priori }probability $\eta _{i\text{.}}$

Now we consider the global fidelity formula Eq. (6). For convenience, let $%
a_{ij}=\left| a_{ij}\right| e^{i\sigma _{ij}}$. If we assume that the
elements of matrix $\Xi $ are real, in order to make the global fidelity
maximal, factors $e^{i\sigma _{ij}}$ and $e^{i\sigma _{ik}}$ must be the
same. So we can write the Eq. (3.1) in a new form. 
\begin{eqnarray}
\left| \Phi _1\right\rangle &=&e^{i\sigma _1}(a_{11}\left| \psi
_1^N\right\rangle +a_{12}\left| \psi _2^N\right\rangle +\cdots +a_{1n}\left|
\psi _n^N\right\rangle )  \eqnum{14} \\
\left| \Phi _2\right\rangle &=&e^{i\sigma _2}(a_{21}\left| \psi
_1^N\right\rangle +a_{22}\left| \psi _2^N\right\rangle +\cdots +a_{2n}\left|
\psi _n^N\right\rangle )  \nonumber \\
&&\vdots  \nonumber \\
\left| \Phi _n\right\rangle &=&e^{i\sigma _n}(a_{n1}\left| \psi
_1^N\right\rangle +a_{n2}\left| \psi _2^N\right\rangle +\cdots +a_{nn}\left|
\psi _n^N\right\rangle ),  \nonumber
\end{eqnarray}
where $a_{ij}$ are real numbers. So it is sufficient to consider the real
number coefficients to find the maximum of the global fidelity. In the next
section we only study the global fidelity in this sense.

\section{Some bounds for state-dependent cloning when state-set contains
three states}

Now we consider the situation that the state set contains three states $%
\{\psi _1,\psi _2,\psi _3\}$. We assume that the three states are taken from
state-set $\{\sin \theta \left| 1\right\rangle +\cos \theta \left|
0\right\rangle ,$ $0\leq \theta \leq \frac \pi 2\}.$ The elements of the
matrix $\Xi $ are naturally real. We consider the quantum cloning process $%
M\rightarrow N$. The quantum state in the space spanned by $\{\left| \psi
_1^N\right\rangle ,\left| \psi _2^N\right\rangle ,\left| \psi
_3^N\right\rangle \}$ is a point on the complex spherical surface (in
general, the states $\left| \psi _1^N\right\rangle ,\left| \psi
_2^N\right\rangle ,\left| \psi _3^N\right\rangle $ are linearly independent
and can span a 3-dimensional space). With the reason pointed out before,
when considering the optimal cloning strategy, we can only consider the
states which have real coefficients, and these states span the spherical
surface $S^2$. Finding the optimal clone is equal to finding three points on
the $S^2$ which make the distances between them and the idea copies minimal.
This situation is described in Fig. 1. In this figure, the edge of the outer
triangle $a^{\text{ }}$corresponds the angle between $\left| \psi
_2\right\rangle ^N$ and $\left| \psi _3\right\rangle ^N$, that is, $\cos a^{%
\text{ }}=(\left\langle \psi _2\right. \left| \psi _3\right\rangle )^N$. The
edge of the inner triangle $a^{^{\prime }}$ corresponds the angle between $%
\left| \psi _2\right\rangle ^M$ and $\left| \psi _3\right\rangle ^M$, that
is, $\cos a^{^{\prime }\text{ }}=(\left\langle \psi _2\right. \left| \psi
_3\right\rangle )^M$. And so on.

In order to get the optimal approximate of the global fidelity, we must give
some characters of the spherical surface $S^2$.

{\it Lemma 2. }For the triangle on the spherical surface, there is a
fundamental formula\cite{book} 
\[
\cos a=\cos b\cos c+\sin a\sin b\cos \alpha , 
\]
where $a,b,c$ are the length of the three edges of this triangle and $\alpha 
$ is the angle between edge $b$ and $c$.

From lemma 2, the following equations can be obtained from Fig. 1 (We
suppose $\alpha =\angle BAC,$ $\theta =\angle BAA^{^{\prime }},$ $\Phi
_1=\angle AA^{\prime }B,$ $\varphi =\angle AA^{^{\prime }}C,$ $\Phi =\angle
AA^{^{\prime }}B^{^{\prime }}$ $,\beta =\angle C^{^{\prime }}A^{^{\prime
}}B^{^{\prime }}$and $c+l\leq \frac \pi 2,b+l\leq \frac \pi 2$) 
\begin{equation}
\cos m=\cos c\cos l+\sin c\sin l\cos \theta ,  \eqnum{15-1}
\end{equation}
\begin{equation}
\frac{\sin c}{\sin \Phi _1}=\frac{\sin m}{\sin \theta }  \eqnum{15-2}
\end{equation}
and 
\begin{equation}
\cos n=\cos b\cos l+\sin b\sin l\cos (\alpha -\theta ),  \eqnum{16-1}
\end{equation}
\begin{equation}
\frac{\sin b}{\sin \varphi }=\frac{\sin n}{\sin \left( \alpha -\theta
\right) }.  \eqnum{16-2}
\end{equation}
Now we can calculate out that 
\begin{eqnarray}
\cos l_2 &=&\cos m\cos c^{^{\prime }}+\sin m\sin c^{^{\prime }}\cos (\Phi
-\Phi _1)  \eqnum{17} \\
&\leq &\cos m\cos c^{^{\prime }}+\sin m\sin c^{^{\prime }}  \nonumber \\
&\leq &\cos m\cos c^{^{\prime }}+\sin (c+l)\sin c^{^{\prime }}.  \nonumber
\end{eqnarray}
We have already used the condition $c+l\leq \frac \pi 2$ , and inequalities $%
m\leq c+l$ (it is proven in the following), that is, $\sin m\leq \sin (c+l)$%
. Substituting Eq. (14-1) into Eq. (16) and rearranging it in order of $\cos
l$ and $\sin l$, we get 
\begin{equation}
\cos l_2\leq \cos (c-c^{^{\prime }})\cos l+(\sin c\cos c^{^{\prime }}\cos
\theta +\cos c\sin c^{^{\prime }})\sin l.  \eqnum{18}
\end{equation}

For the same reason, we can get 
\begin{eqnarray}
\cos l_3 &=&\cos n\cos b^{^{\prime }}+\sin n\sin b^{^{\prime }}\cos (2\pi
-\Phi -\beta -\varphi )  \eqnum{19} \\
&\leq &\cos n\cos b^{^{\prime }}+\sin n\sin b^{^{\prime }}  \nonumber \\
&\leq &\cos n\cos b^{^{\prime }}+\sin (b+l)\sin b^{^{\prime }}.  \nonumber
\end{eqnarray}

Now inserting Eq.(15-1) into this formula and using the fact that $\sin
\theta \leq \sin \alpha ,$ we get 
\begin{eqnarray}
\cos l_3 &\leq &[\cos b\cos l+\sin b\sin l\cos (\alpha -\theta )]\cos
b^{^{\prime }}+\sin (b+l)\sin b^{^{\prime }}  \eqnum{20} \\
&\leq &\cos (b-b^{^{\prime }})\cos l+[\sin b\cos b^{^{\prime }}\cos \alpha
\cos \theta +(\sin b\cos b^{^{\prime }}\sin ^2\alpha +\cos b\sin b^{^{\prime
}})]\sin l  \nonumber
\end{eqnarray}

From the definition of the global fidelity, we insert Eq. (14) and Eq. (15)
into the fidelity formula to get 
\begin{eqnarray}
F_g &=&\frac 13(\cos ^2l+\cos ^2l_2+\cos ^2l_3)  \eqnum{21} \\
&\leq &\frac 13(\cos ^2l+[\cos (c-c^{^{\prime }})\cos l+(\sin c\cos
c^{^{\prime }}\cos \theta +\cos c\sin c^{^{\prime }})\sin l]^2  \nonumber \\
&&+[\cos (b-b^{^{\prime }})\cos l+[\sin b\cos b^{^{\prime }}\cos \alpha \cos
\theta +(\sin b\cos b^{^{\prime }}\sin ^2\alpha +\cos b\sin b^{^{\prime
}})]\sin l]^2).  \nonumber
\end{eqnarray}

So the upper bound of the fidelity must be less than the maximum of the
right hand of Eq.(16). Before getting the result, let $A_1=\cos
(c-c^{^{\prime }}),$ $A_2=\sin c\sin c^{^{\prime }},$ $A_3=\cos c\sin
c^{^{\prime }};$ $B_1=\cos (b-b^{^{\prime }}),$ $B_2=\sin b\cos b^{^{\prime
}}\cos \alpha ,$ $B_3=\sin b\cos b^{^{\prime }}\sin ^2\alpha +\cos b\sin
b^{^{\prime }}.$ Then the maximum of the right hand of Eq. (16) is 
\begin{equation}
\frac 13(\cos ^2l+[A_1\cos l+(A_2+A_3)\sin l]^2+[B_1\cos l+(B_2+B_3)\sin
l]^2),  \eqnum{22}
\end{equation}
where $l$ satisfies the condition 
\begin{equation}
tg(2l)=\frac{2[A_1(A_2+A_3)+B_1(B_2+B_3)}{%
1+A_1^2+B_2^2-(A_2+A_3)^2-(B_2+B_3)^2}.  \eqnum{23}
\end{equation}
It can be seen from these formulas that they are symmetric for $A$ and $B$.

We can find another interesting thing from this spherical surface $S^2.$ We
can attain all of the results of Rastegin\cite{r0108,0111} succinctly and
directly from Cauchy Lemma , which has obvious geometric meaning. The Cauchy
Lemma on the $S^2$ is very important and useful. It is given out as the
following.

{\it Lemma 3.} ({\it Cauchy }Lemma{\it ) }There are two polygons $%
A_1A_2\cdots A_n$ and $B_1B_2\cdots B_n$. If the lengths of the edges in
these two polygons satisfy $A_1A_2=B_1B_2$, $A_2A_3=B_2B_3$, $\cdots $, $%
A_{n-1}A_n=B_{n-1}B_n$ and the angles satisfy $A_2\leq B_2$, $A_3\leq B_3$, $%
\cdots $, $A_{n-1}\leq B_{n-1}$. Then $A_1A_n\leq B_1B_n.$

This lemma looks very simple, but its proof is rather difficult, the proof
of this lemma can be found in\cite{book}. This Lemma is useful to get some
inequality. When $n=3$, we get the familiar inequality for spherical surface
triangle 
\[
A_1A_3-A_3A_2\leq A_1A_2\leq A_1A_3+A_3A_2. 
\]
The left part of this formula is the edge of a special triangle $A_1A_3A_2$
whose angle $A_3$ is $0$. While any angle of a triangle must be not less
than $0$, the left inequality is held. For the same reason, the right
inequality is also hold. This inequality is just the same as $\cos \delta
_{\Phi \Psi }\leq \cos (\delta _{\Phi \Upsilon }-\delta _{\Upsilon \Psi })$
which was introduced by Rastegin\cite{r0108}. When $n=4$ we can get some
useful inequality (Fig. 1) 
\begin{equation}
l+c^{^{\prime }}+l_2\geq c,\text{ }l+b^{^{\prime }}+l_3\geq b,\text{ }%
l_2+a^{^{\prime }}+l_3\geq a.  \eqnum{24}
\end{equation}

This inequality is the same as $\cos (\delta _{\Gamma \Lambda }+\delta
_{\Gamma \Xi }+\delta _{\Lambda \Sigma })\leq \cos \delta _{\Xi \Sigma }$ $%
\left( \delta _{\Gamma \Lambda }+\delta _{\Gamma \Xi }+\delta _{\Lambda
\Sigma }\leq \frac \pi 2\right) $ in the Rastegin's paper\cite{r0108}. Using
Cauchy lemma, we can make the condition weaken to $\delta _{\Gamma \Lambda
}+\delta _{\Gamma \Xi }\leq \frac \pi 2$, and get the useful new inequality $%
\cos (\delta _{\Gamma \Lambda }+\delta _{\Gamma \Xi })\leq \cos (\delta
_{\Xi \Sigma }-\delta _{\Lambda \Sigma })$. From these new inequalities we
can get a new upper bound of the three states global fidelity 
\begin{eqnarray}
F_g &=&\frac 13(\cos ^2l+\cos ^2l_2+\cos ^2l_3)  \eqnum{25} \\
&\leq &\frac 16(3+\cos (l+l_1)+\cos (l_1+l_2)+\cos (l_2+l_3))  \nonumber \\
&\leq &\frac 16(3+\cos (a-a^{^{\prime }})+\cos (b-b^{^{\prime }})+\cos
(c-c^{^{\prime }})).  \nonumber
\end{eqnarray}
This bound of the global fidelity is symmetric to the three edges of the
triangle, that is, symmetric to the three states. This formula is more
simple than Eq.(17). The equal sign is held when the states in the state-set
are orthogonal.

\section{Some bounds of state-dependent cloning when state-set contains N
states}

When the state-set contains more than three states, they can span a space
more than three dimensions. In this situation, the actual quantum cloned
states are points on a spherical surface more than $\ 2$-dimension. The
method which we used to get Eq. (17) on $S^2$ is not available. Fortunately,
the inequality (19) for any four states on the same spherical surface is
still correct. So we can use this inequality to get some upper bounds of the
global fidelity of multi-state-dependent quantum cloning.

Assume the state-set contains $n$ states $\{\psi _1,\psi _2,\cdots ,\psi
_n\} $. And these states are taken from $\{\sin \theta \left| 1\right\rangle
+\cos \theta \left| 0\right\rangle ,$ $0\leq \theta \leq \frac \pi 2\}$. At
first, we divide the state-set $\{\psi _1^N,\psi _2^N,\cdots ,\psi _n^N\}$
into several groups $\{\{\psi _1^N,\psi _2^N,\cdots ,\psi _i^N\}$, $\{\psi
_{i+1}^N,\psi _{i+2}^N,\cdots ,\psi _j^N\}$, $\cdots $, $\{\psi
_{k+1}^N,\psi _{k+2}^N,\cdots ,\psi _n^N\}\}$, and every group constitutes a
convex polygon on the same $S^2$ (We can obtain this result by the following
step. At first, we can take any three states from the state-set and they
must constitute the spherical surface $S^2$. Then put all of the vectors
which are linearly dependent on the three states and make the points of all
of these vectors constitute a convex polygon. Do the same operation to the
rest vectors until the number of the vectors is less than 3. At this
situation, we take some vectors from the group which has more than three
vectors to get a new spherical surface $S^2$). On the spherical surface $S^2$%
, we can use the Cauchy lemma to get the inequality (19) and insert them
into the formula of the global fidelity. 
\begin{eqnarray}
F_g &=&\frac 1n[\sum_{p=1}^i\cos ^2l_p+\sum_{q=i+1}^j\cos ^2l_q+\cdots
+\sum_{r=k+1}^n\cos ^2l_{,r}]  \eqnum{26} \\
&\leq &\frac 1{2n}[(i+\sum_{p=1}^i\cos (a_{p+1,p}-a_{p+1,p}^{^{\prime
}}))+\sum_{q=i+1}^j((j-i)+\cos (a_{q+1,q}-a_{q+1,q}^{^{\prime }}))+\cdots 
\nonumber \\
&&+\sum_{r=k+1}^n((n-k)+\cos (a_{r+1,r}-a_{r+1,r}^{^{\prime }})]  \nonumber
\\
&=&\frac 12+\frac 1{2n}[\sum_{p=1}^i\cos (a_{p+1,p}-a_{p+1,p}^{^{\prime
}})+\sum_{q=i+1}^j\cos (a_{q+1,q}-a_{q+1,q}^{^{\prime }})+\cdots  \nonumber
\\
&&+\sum_{r=k+1}^n\cos (a_{r+1,r}-a_{r+1,r}^{^{\prime }})],  \nonumber
\end{eqnarray}

where $\cos l_i=\left\langle \Phi _i\right. \left| \psi _i^N\right\rangle $
and $\cos a_{i+1,i}=\left\langle \psi _{i+1}^N\right. \left| \psi
_i^N\right\rangle $, $\cos a_{i+1,i}^{^{\prime }}=\left\langle \psi
_{i+1}^M\right. \left| \psi _i^M\right\rangle $ and $1,2,\cdots i$; $%
i+1,i+2,\cdots ,j$; $\;\cdots $; $k+1,k+2,\cdots n$ are vertexes of a convex
polygon on $S^2$ respectively.

It can be seen from the deriving process that the result is dependent on the
partition of the states and the choice of the loops. It is not good enough
for us since it is not uniquely determined by the state-set. Since the Eq.
(19) is correct for every four vectors, we can get a more symmetric result.
We can average all of the possible divided sets and get 
\begin{equation}
F_g\leq \frac 12+\frac 1{2n(n-1)}[\sum_{k\neq j=1}^n\cos
(a_{j,k}-a_{j,k}^{^{\prime }})].  \eqnum{27}
\end{equation}

Even though this inequality is the most symmetry for every state-set, it is
not the tightest bound which we can get by this method for the quantum
cloning process. There are some methods to refine the bound. First, we
construct the $n\times n$ matrix $M$ whose elements are $a_{j,k}-a_{j,k}^{^{%
\prime }}$, So the matrix is 
\begin{equation}
M=\left( 
\begin{array}{llll}
a_{1,1}-a_{1,1}^{^{\prime }} & a_{1,2}-a_{1,2}^{^{\prime }} & \cdots & 
a_{1,n}-a_{1,n}^{^{\prime }} \\ 
a_{2,1}-a_{2,1}^{^{\prime }} & a_{2,2}-a_{2,2}^{^{\prime }} & \cdots & \vdots
\\ 
\vdots & \vdots & \ddots & \vdots \\ 
a_{n,1}-a_{n,1}^{^{\prime }} & \cdots & \cdots & a_{n,n}-a_{n,n}^{^{\prime }}
\end{array}
\right) .  \eqnum{28}
\end{equation}

It should be pointed out that the diagonal elements of this matrix are zero.
The first step of our refining process is to find the maximal element $%
M_{mn}(m>n)$ in the upper triangle of the matrix which is denoted by $e_1$.
Then we replace the row and the column in which the element $e_1$ lies by
zeroes. At the end of the first step we get an element and a new $n\times n$
matrix. (If there are $i$ maximal elements which are denoted by $M_{l_1k_1},$
$M_{l_2k_2}\cdots $ $M_{l_ik_i}$ in the upper triangle of the matrix $M$.
Then we denote the secondary maximum element of upper triangle elements
within the rows $l_{j\text{ }}$and columns $k_j$ as $M_{l_j^{^{\prime
}}k_j^{\prime }}$. Thus we denote the minimal element $M_{l_{j0}^{^{\prime
}}k_{j0}^{^{\prime }}}$ of $M_{l_j^{^{\prime }}k_j^{^{\prime }}}$ as $e_1$).
We can iterate the operation to get the elements $e_3$ ,$e_4$ ,$\cdots ,e_{[%
\frac{n+1}2]}$($[x]$ is the integer part of $x$). Then a more stringent
bound can be attained with our method 
\begin{equation}
F_g\leq \frac 12+\frac 1{2[\frac{n+1}2]}[\sum_{j=1}^{[\frac{n+1}2]}\cos e_j].
\eqnum{29}
\end{equation}

\section{Conclusion}

In the practical quantum information processes the state-dependent cloning
is more important than the universal cloning. In fact, we can view the
universal cloning as the lower bound of the state-dependent cloning. If we
have no information about the state-set, the optimal strategy we can chose
is the universal cloning. But if we know something about the state-set, that
is, we have some information about the state, we can find the strategy no
worse than the universal cloning. In fact, we always have some information
about the states in the practical information processes. So it is very
important to find the strategy of the state-dependent cloning which is
better than universal clone. At the same time we want to know {\it how much}
we can improve the quantum cloning process when we know something about the
state-set. Unfortunately, it is too difficult to solve this problem
completely. We can only find some bounds for this process and partly answer
this question.

There are close relations between state-dependent cloning and eavesdropping
in quantum cryptography. Cloning is a method for eavesdropper to
eavesdropping (but it is not necessarily the optimal one). When the
state-set only contains two states, the situation has already been
completely discussed by Bru\ss\ {\it et al.} \cite{brub98}. Because of the
relationship between the quantum cloning and eavesdropping, the bound of the
multi-state-dependent quantum cloning can be considered as a bound for the
eavesdropping when the eavesdropper use the cloning strategy to get the
information of the communication process.

\section{Acknowledgment}

This work was funded by the National Fundamental Research Program
(2001CB309300), the Innovation Funds from Chinese Academy of
Sciences.

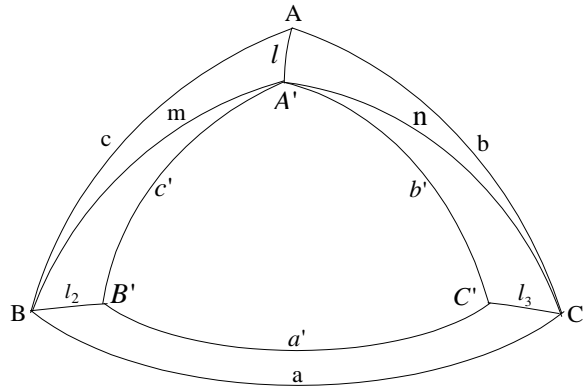
\end{document}